\documentclass{iau}
\usepackage{graphicx}
\usepackage{url}

\title[IAUS291.~~Origin of pulsar pulse fine structure ] 
{Origin of the pulsar pulse fine structure}

\author[O. M. Ulyanov, A. A. Seredkina \& A. I. Shevtsova ]   
{O. M. Ulyanov,
A. A. Seredkina \and
A. I. Shevtsova}

\affiliation{Department of Astrophysics, Institute of Radio Astronomy of NAS of Ukraine, \\ Krasnoznamennaya str.4,
Kharkov 61002, Ukraine\\ email: {\tt oulyanov@rian.kharkov.ua, seredkina.a@gmail.com, alice.shevtsova@gmail.com} } 

\pubyear{2012}
\volume{291}  
\jname{\mbox{Neutron Stars and Pulsars: Challenges and
Opportunities after 80 years}}
\editors{J. van Leeuwen, ed.}
\begin{document}

\maketitle

\begin{abstract}
We give a new numerical model of pulsar pulse radiation through
the interstellar medium (ISM) considering the propagation effects.
It explains the deficit of a scattering measure at the decameter
range of frequencies that leads to the possibility of detecting
the pulsar pulse fine structure. The results of numerical
simulation confirm that the fine structure may be detected at low
frequencies and this is qualitatively agreed with the
observational data. 
\end{abstract}

\firstsection 
\section{Introduction}

Despite the fact that the micro structure of pulsar radio emission
was discovered more than 40 years ago (\cite[Hankins
1971]{Hankins71}) still there is no generally accepted model of an
origin of this phenomenon. We consider a conception of a fine
structure of pulsar radio emission that covers different time
scales from nanoseconds in the centimeter wave range to
milliseconds in the decameter wave range. We propose a new model
of the fine structure formation that is considered as a result of
the propagation of the radio pulses through the interstellar
medium and subsequent processing of the received signal in the lab
frame.

\section{Observational Data}
\begin{figure}[b]
\begin{center}
 \includegraphics[width=\textwidth]{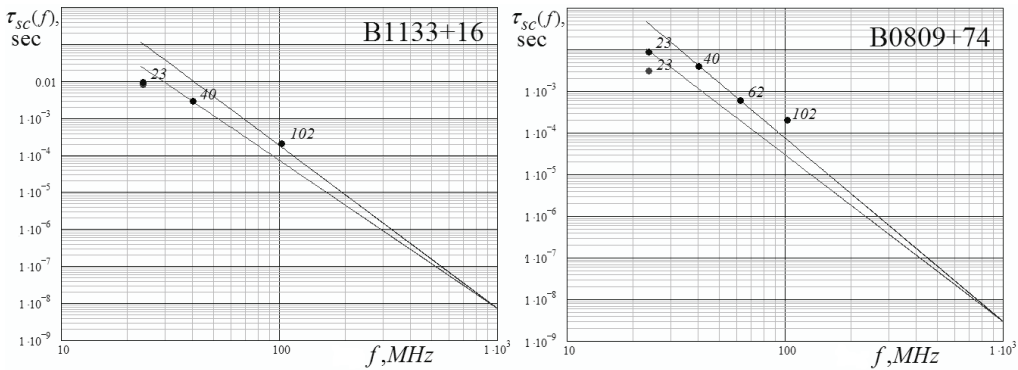}
 \caption{Scattering time constant at different frequencies.
 The normal spectrum of electron inhomogeneities (upper line);
 Kolmogorov spectrum of electron inhomogeneities (lower line).}
   \label{fig1}
\end{center}
\end{figure}
The model gives us the value of the scattering time constant $
\tau_{sc}( f )$ at the frequency $f$  that describes the temporal
broadening of pulses due to interstellar scattering: $ \tau_{sc}
(f) = \tau_{sc,0} (f_0) {\left (  {f}/ {f_0} \right )} ^{-\alpha},
$
where $\tau_{sc,0} (f_0)$ is the scattering time constant at the
fixed frequency $f_0$ that one could take from a pulsar catalogue
(usually $f_0$ = 1GHz); $\alpha$ is a power that corresponds to
different spectra of spatial inhomogeneities of the electron
concentration. Thus $\alpha$ = 4.0 for the normal spectrum of the
electron density fluctuations, $\alpha$ = 4.4 of the Kolmogorov
spectrum.

The observation in the decameter range (\cite[Ul'yanov \&
Zakharenko 2012]{Ul'yanovZakharenko06}) gives us the lower value
of the scattering measure for anomalous-intensity pulses
(Fig.\,\ref{fig1}) . It means that the modern model of the radio
pulses propagation through the ISM is not completely correct. On
the other hand, we are able to detect the fine structure of the
pulsar pulses in the low frequency range that gives us more
possibilities to study this phenomenon. The purpose of the present
work is to create a model of ISM to explain the deficit of a
scattering measure and formation of the fine structure of the
pulsar radio emission.

\section{Numerical Simulation}

The short radio pulses have been generated in a wide frequency
range with the pulsar rotation period P = 1 sec. The pulses are
represented as an amplitude modulated noise signal that has a
Gauss shape envelope. The width of the envelope is about 10\% of
the pulses period. The signal has the uniform distribution of the
frequencies in the receiving range $\Delta f=f_H - f_L$ and the
uniform distribution of the initial phases in the angles range
$[-\pi,\pi]$. The noise signal amplitudes have the Rayleigh
distribution. Also the white noise with the normal distribution is
added to the propagation channel.

The main idea of the present paper is to create an ISM model 
that describes the scattering measure deficit. We are
taking into account two main factors: the dispersion
delay of low frequencies versus high frequencies and the
scattering by space irregularities of  the electron
concentration. We do not consider the Faraday rotation effect.

The dispersion delay of low frequencies versus high frequencies is
the result of the relationshep between the refraction index and frequency in the cold anisotropic ISM plasma. Using
the eikonal equation we write the delay phase $\varphi (\omega)$:
\begin{equation}
{\varphi (\omega)} \approx \omega \frac{L}{c} -
\frac{1}{\omega} \frac{2 \pi e^2}{m_e c} \int_0^L N_e(z)d z ,
\end{equation}
where $\omega$ is a cyclic frequency; $e$, $m_e$  are the electron
charge and its rest mass; $c$ is the speed of light. $DM=\int_0^L N_e(z)d z$ is the dispersion measure, which characterizes the number of free
electrons along the line of sight and is measured in units of
pc$\cdot$cm$^{-3}$.

We can also define the time delay of a frequency $\omega_i$ versus
infinity to be able to reduce the dispersion in the received
signal . From the equation (3.1):
 $ \Delta \tau_{DM} (DM, \omega_i) = \frac{2 \pi e^2}{m_e c} DM {\left ( \frac{1} {\omega_i} \right )}^2  $

We also consider the scattering of the pulsar radiation by
fluctuations in the concentration of free electrons in the
propagation medium along the line of sight. These lead to the
amplitude and phase fluctuations of the received wave, 
called scintillations. To describe the scintillations we use the
thin screen model. In this model we consider only phase
fluctuations of the original signal that occur in the thin screen
that is perpendicular to the line of sight, located halfway
the source and observer. The distance between
the screen and the observer or the screen and the source must be
much greater than the signal wavelength.  In this case one could
use the eikonal equation. Scattering on the thin screen leads to
multibeam interference at 
the receiver and the point source becomes a finite extended object
with angular size equal to the scattering angle $\theta_{sc}$.

The geometric optic approximation gives the scattering angle as a
function $\theta_{sc}(\lambda)$ of wavelength: $ \theta_{sc}
(\lambda) \approx \frac{\Delta \varphi(\lambda)}{2\pi}
\frac{\lambda}{a} = \frac{r_e \delta N_e \lambda^2}{2\pi}
(\frac{DW}  {a})^{1/2}$  , where $\lambda$ is the wavelength,
$r_e$ is the electron radius, $\delta N_e$ is the fluctuation of
the electron concentration, $a$ is the characteristic length scale
of the electron-concentration irregularities, and $DW$ is
the screen width.

At the receiver the signal passes through a preselector filter.
Then, we used the method of dispersion delay removing proposed by
\cite{HankinsRickett75}. The method uses the eikonal equation to
find the argument in the transfer function of the ISM. For the
narrow band signal this argument $\Delta \alpha_T (\omega)$ can be
expanded in a Taylor series centered at the receiver central
frequency $f_0$. The initial phase in the observational reference
frame can be written as:
$
\Delta \alpha_T (\omega)= \frac{2\pi e^2} {m_e c} DM  \left ({
\frac{1} {\omega_i}} - {\frac{ \Delta \omega} {{\omega_i}^2} } +
{\frac{{\Delta \omega}^2} {\omega_i^3} } - {\frac{{\Delta
\omega}^3}{{\omega_i}^4}} + ...\right )
$

\section{Main Results}
\begin{figure}[t]
\begin{center}
 \includegraphics[width=5in]{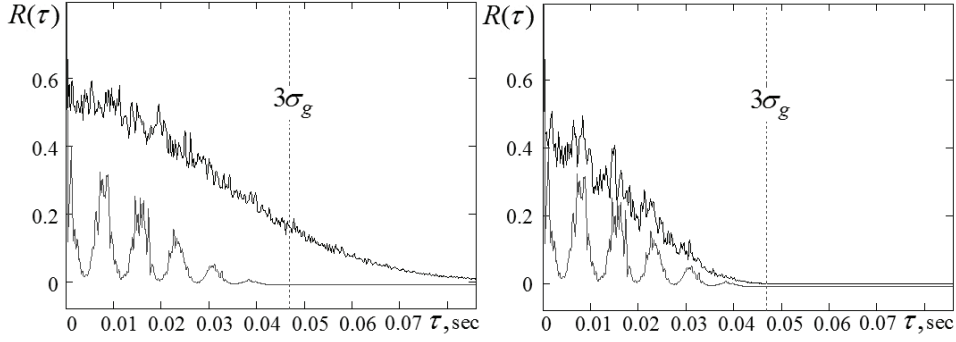}
 \caption{Input (lower) and output (upper) ACFs of the single pulses with the fine structure.
  Large-scale and short-scale irregularities are given on the left and right sides respectively.}
   \label{fig2}
\end{center}
\end{figure}

As a result of the numerical simulation we obtained signal
responses at different stages of the modeling. Then we studied the
signals by using ACF and spectral analysis. At the sky frequency
$f_0=20$\,MHz the fine structure of  the model pulse may be or may not be
detected with different parameters of the ISM. The main parameter
is the  spatial irregularities size of the electron density. The
data analysis results are shown in Fig.\,\ref{fig2}. It
is qualitatively similar to real observational data\footnote{For more details see \\\url{www.pulsarastronomy.net/IAUS291/download/Posters/IAUS291_UlyanovO_257.pdf}}.

\section{Conclusions}

1)   The presence or absence of the fine structure in the low
frequency range is explained by a reaction of radio waves on the
propagation in a plasma environment, which is located in the line
of sight.

2)   The fine structure is smoothed out more strongly by
scattering at the large-scale spatial electron density
irregularities, at the same time one can detect the fine structure
of the pulsar radiation in case of scattering by small-scale
irregularities.

3)  The characteristic width of the detected fine structure is
frequency dependent and increases with decreasing of the radiation
frequency.

\end{document}